\documentclass[twocolumn]{aastex631}
\usepackage{graphicx,xspace,apjfonts}

\newcommand\Peg{51\,Peg\xspace}
\newcommand\Pegb{51\,Peg\,b\xspace}
\newcommand\HD{HD\,76151\xspace}
\newcommand\Sco{18\,Sco\xspace}
\newcommand\Cyg{16\,Cyg\xspace}
\newcommand\CygA{16\,Cyg\,A\xspace}
\newcommand\xray{\mbox{X-ray}\xspace}

\shorttitle{Weakened Magnetic Braking in 51\,Peg}
\shortauthors{Metcalfe et al.}

\journalinfo{The Astrophysical Journal Letters}

\begin{document}

\title{\Large Weakened Magnetic Braking in the Exoplanet Host Star 51\,Peg}

\author[0000-0003-4034-0416]{Travis S.~Metcalfe}
\affiliation{White Dwarf Research Corporation, 9020 Brumm Trail, Golden, CO 80403, USA}

\author[0000-0002-6192-6494]{Klaus G.~Strassmeier} 
\affiliation{Leibniz-Institut f\"ur Astrophysik Potsdam (AIP), An der Sternwarte 16, D-14482 Potsdam, Germany}

\author[0000-0002-0551-046X]{Ilya V.~Ilyin} 
\affiliation{Leibniz-Institut f\"ur Astrophysik Potsdam (AIP), An der Sternwarte 16, D-14482 Potsdam, Germany}

\author[0000-0002-1988-143X]{Derek Buzasi}
\affiliation{Department of Chemistry and Physics, Florida Gulf Coast University, 10501 FGCU Blvd S, Fort Myers, FL 33965, USA}

\author[0000-0003-3061-4591]{Oleg Kochukhov} 
\affiliation{Department of Physics and Astronomy, Uppsala University, Box 516, SE-75120 Uppsala, Sweden}

\author[0000-0002-1242-5124]{Thomas R.~Ayres} 
\affiliation{Center for Astrophysics and Space Astronomy, 389 UCB, University of Colorado, Boulder, CO 80309, USA}

\author[0000-0002-6163-3472]{Sarbani Basu}
\affiliation{Department of Astronomy, Yale University, PO Box 208101, New Haven, CT 06520-8101, USA}

\author[0000-0003-1125-2564]{Ashley Chontos}
\affiliation{Department of Astrophysical Sciences, Princeton University, Princeton, NJ 08544, USA}

\author[0000-0002-3020-9409]{Adam J.~Finley} 
\affiliation{Department of Astrophysics-AIM, University of Paris-Saclay and University of Paris, CEA, CNRS, Gif-sur-Yvette Cedex F-91191, France}

\author[0000-0001-5986-3423]{Victor See} 
\affiliation{European Space Agency (ESA), European Space Research and Technology Centre (ESTEC), Keplerlaan 1, 2201 AZ Noordwijk, the Netherlands}

\author[0000-0002-3481-9052]{Keivan G.~Stassun} 
\affiliation{Vanderbilt University, Department of Physics \& Astronomy, 6301 Stevenson Center Lane, Nashville, TN 37235, USA}

\author[0000-0002-4284-8638]{Jennifer L.~van~Saders} 
\affiliation{Institute for Astronomy, University of Hawai`i, 2680 Woodlawn Drive, Honolulu, HI 96822, USA}

\author[0000-0002-8621-2682]{Aldo G.~Sepulveda}
\altaffiliation{NSF Graduate Research Fellow}
\affiliation{Institute for Astronomy, University of Hawai`i, 2680 Woodlawn Drive, Honolulu, HI 96822, USA}

\author[0000-0003-2058-6662]{George R.~Ricker} 
\affiliation{Department of Physics and Kavli Institute for Astrophysics and Space Science, Massachusetts Institute of Technology, Cambridge, MA 02139, USA}

\begin{abstract}

The consistently low activity level of the old solar analog 51\,Peg not only facilitated 
the discovery of the first hot Jupiter, but also led to the suggestion that the star 
could be experiencing a magnetic grand minimum. However, the 50 year time series showing 
minimal chromospheric variability could also be associated with the onset of weakened 
magnetic braking (WMB), where sufficiently slow rotation disrupts cycling activity and 
the production of large-scale magnetic fields by the stellar dynamo, thereby shrinking 
the Alfv\'en radius and inhibiting the efficient loss of angular momentum to magnetized 
stellar winds. In this Letter, we evaluate the magnetic evolutionary state of 51\,Peg by 
estimating its wind braking torque. We use new spectropolarimetric measurements from the 
Large Binocular Telescope to reconstruct the large-scale magnetic morphology, we 
reanalyze archival X-ray measurements to estimate the mass-loss rate, and we detect 
solar-like oscillations in photometry from the Transiting Exoplanet Survey Satellite, 
yielding precise stellar properties from asteroseismology. Our estimate of the wind 
braking torque for 51\,Peg clearly places it in the WMB regime, driven by changes in the 
mass-loss rate and the magnetic field strength and morphology that substantially exceed 
theoretical expectations. Although our revised stellar properties have minimal 
consequences for the characterization of the exoplanet, they have interesting 
implications for the current space weather environment of the system.

\end{abstract}


\section{Introduction}\label{sec1}

Decades before the first hot Jupiter was discovered orbiting the old solar analog \Peg 
\citep{Mayor1995}, long-term monitoring of its chromospheric activity began at the Mount 
Wilson Observatory \citep[MWO;][]{Wilson1968}. These observations revealed nearly 
constant activity \citep[$\log R'_{\rm HK}=-5.068$;][]{Henry2000}, below the solar 
minimum level \citep[$\log R'_{\rm HK}=-4.984$;][]{Egeland2017}, starting in 1966 and 
continuing to the present day \citep{Baliunas1995, Radick2018, Baum2022}. Despite the low 
level of chromospheric activity, several seasons of the MWO data showed variability 
attributed to stellar rotation, with periods ranging from 21.3--22.6~days 
\citep{Henry2000}. A reanalysis of these same measurements confirmed a rotation period 
$P_{\rm rot} = 21.9 \pm 0.4$~days from the observations in 1998 \citep{Simpson2010}, and 
highlighted a disagreement between the observed rotation period and the value predicted 
from the mean activity level \citep[29~days;][]{Wright2004}.

 \begin{figure*}[t]
 \centering\includegraphics[height=\textwidth,angle=90]{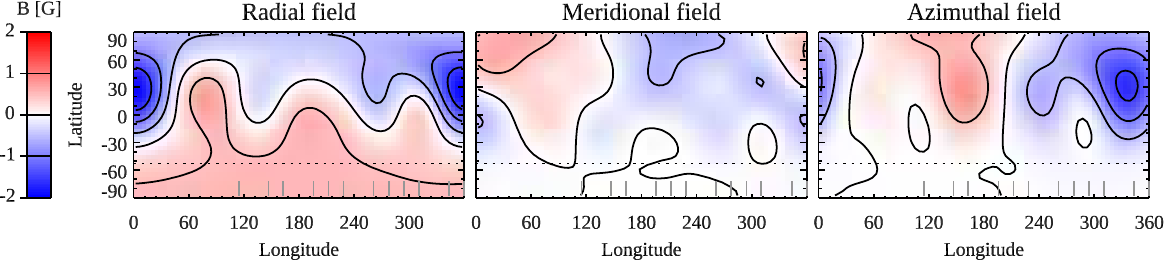}
 \caption{ZDI maps of the radial, meridional, and azimuthal field components of \Peg. 
Contours are shown with a step of 0.5~G. The dotted line corresponds to the lowest 
visible latitude. The vertical bars at the bottom of each panel show the central 
longitude of each LBT observation.\\ \label{fig1}}
 \end{figure*} 

The consistently low activity level, also seen in \xray measurements, led to the 
suggestion of \Peg as a candidate Maunder minimum star \citep{Poppenhager2009}. The 
Maunder minimum was the 70 year interval between 1645 and 1715 when very few sunspots 
appeared on the solar disk, and the phenomenon is more generally known as a magnetic 
grand minimum \citep{Usoskin2007}. The absence of long-term chromospheric variations in a 
50 year time series cannot determine whether a star has temporarily or permanently lost 
its activity cycle. The only unambiguous evidence that a star has experienced a magnetic 
grand minimum is the observation of a transition from cycling to non-cycling or from 
non-cycling to cycling. Currently, such evidence only exists for one star, HD\,166620 
\citep{Baum2022, Luhn2022}. Another interpretation of constant activity stars like \Peg 
was put forward by \cite{Metcalfe2017}, who suggested that they may represent the 
disappearance of activity cycles associated with the onset of weakened magnetic braking 
\citep[WMB;][]{vanSaders2016, vanSaders2019, Hall2021}. In this scenario, sufficiently 
slow rotation disrupts cycling activity and the production of large-scale magnetic fields 
by the stellar dynamo, thereby shrinking the Alfv\'en radius and inhibiting the efficient 
loss of angular momentum to magnetized stellar winds \citep{Metcalfe2022}.

In this Letter, we evaluate the magnetic evolutionary state of \Peg by estimating its 
wind braking torque using the prescription of \cite{FinleyMatt2018}. In 
Section~\ref{sec2.1} we use new spectropolarimetric measurements from the 2$\times$8.4\,m 
Large Binocular Telescope (LBT) to reconstruct the large-scale magnetic morphology of 
\Peg. In Section~\ref{sec2.2} we use archival \xray measurements to estimate the 
mass-loss rate from the empirical relation of \cite{Wood2021}. In Section~\ref{sec2.3} we 
use a detection of solar-like oscillations from the Transiting Exoplanet Survey Satellite 
\citep[TESS;][]{Ricker2014} to place constraints on the stellar radius, mass, and age. In 
Section~\ref{sec3} we bring the measurements together to estimate the wind braking torque 
for \Peg, and in Section~\ref{sec4} we discuss the implications of WMB on the space 
weather environment of this iconic planetary system.

\section{Stellar Properties}\label{sec2}

\subsection{Spectropolarimetry}\label{sec2.1} 

We observed \Peg from the LBT on 12 nights between 2022 November 18 and 2022 December 3 
using the Potsdam Echelle Polarimetric and Spectroscopic Instrument 
\citep[PEPSI;][]{Strassmeier2015, Strassmeier2018}. The instrument configuration and data 
reduction methods were the same as those described in \cite{Metcalfe2019b}, and we 
derived precise mean intensity and circular polarization (Stokes~$V$) profiles at each 
epoch using the least-squares deconvolution \citep[LSD;][]{Kochukhov2010} technique. We 
did not consider linear polarization (Stokes $Q$ and $U$) because the Zeeman signatures 
are typically an order of magnitude smaller \citep{Kochukhov2011}. The LSD analysis 
employed spectral line data from the VALD database \citep{Ryabchikova2015}, and we 
adopted spectroscopic parameters from \cite{Brewer2016}. The observations spanned 15 
nights, corresponding to central longitudes covering 68\% of the 21.9~day rotation 
period, allowing us to reconstruct the large-scale magnetic field with Zeeman Doppler 
Imaging \citep[ZDI;][]{Kochukhov2016}. Although poor weather prevented additional 
observations that could have provided redundant information for the inversion procedure, 
our set of disk-integrated Stokes~$V$ profiles provide some constraints at all stellar 
longitudes. A complete archive of the reduced data is available at 
\dataset[doi:10.5281/zenodo.8381444]{https://doi.org/10.5281/zenodo.8381444}.

The inclination of the stellar rotation axis was estimated using the analytic expressions 
from \cite{Bowler2023} that approximate the Bayesian framework of \cite{MasudaWinn2020} 
to calculate stellar inclination posteriors given measurements of $v \sin i$, rotation 
period, and radius. The posterior inclination distribution for \Peg (from the 
spectroscopic $v \sin i$, the chromospheric rotation period, and the asteroseismic radius 
in Section~\ref{sec2.3}) peaks at $i = 53^\circ$, with a 68\% credible interval between 
$42^\circ$ and $78^\circ$. The lower bound corresponds to $v \sin i=1.8$~km~s$^{-1}$ 
determined by \cite{Morris2019}, while the upper bound is comparable to the orbital 
inclination of \Pegb inferred by \cite{Birkby2017}. Lower inclinations yield less 
geometric cancellation, and thus weaker inferred fields for a given set of Stokes~$V$ 
profiles. Higher inclinations are disfavored by the absence of transits in the \Peg 
system. We adopt the most probable value of the inclination for our fiducial model, and 
assess the impact of the uncertainty in Section~\ref{sec3}.

\begin{figure}[t]
 \centering\includegraphics[width=0.95\columnwidth]{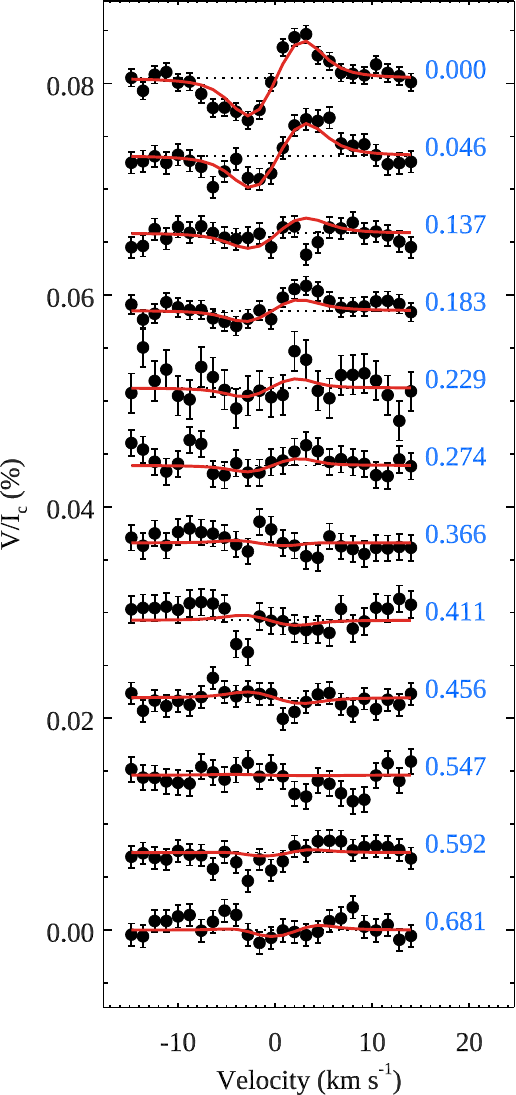}
 \caption{ZDI fit to the observed Stokes~$V$ LSD profiles for \Peg. Observations (points) 
and model profiles (solid lines) are shifted vertically according to the rotational 
phase, which is indicated on the right of each profile.\label{fig2}}
 \end{figure}

Results of the ZDI inversion for \Peg are presented in Figure~\ref{fig1}, which shows 
Mercator maps of the radial, meridional, and azimuthal components of the large-scale 
magnetic field. The vertical bars at the bottom of each panel show the central longitude 
of each LBT observation, and regions of the ZDI map below the dotted lines in 
Figure~\ref{fig1} are not visible from Earth. The magnetic morphology is predominantly 
poloidal (86\% of the field energy is in poloidal components) and nonaxisymmetric 
($m\ne0$ harmonic modes contain 74\% of the field energy). The average field strength is 
$\left<B\right>=0.68$~G with a maximum local strength of 2.28~G. The field structure is 
dominated by a nonaxisymmetric dipole (strength 0.77~G, obliquity 143$^\circ$ towards the 
positive pole), which comprises 64\% of the field energy. The sum of the quadrupole and 
octupole modes contribute 31\% of the field energy. The fit to the observed Stokes~$V$ 
profiles is shown in Figure~\ref{fig2}.

The rate of angular momentum loss due to the magnetized stellar wind depends primarily on 
the radial component of the large-scale field. The prescription of \cite{FinleyMatt2018} 
that we use in Section~\ref{sec3} to estimate the wind braking torque requires polar 
field strengths of the axisymmetric (\mbox{$m=0$}) dipole, quadrupole, and octupole 
components of this field. The observed ZDI map is dominated by nonaxisymmetric 
(\mbox{$m\neq0$}) components, but we follow the procedure described in 
\cite{Metcalfe2022} to calculate the equivalent polar field strengths ($B_{\rm d}, B_{\rm 
q}, B_{\rm o}$) for use with the wind braking prescription. For each spherical harmonic 
degree $\ell$, this procedure calculates the total magnetic flux $\Phi_\ell = \int 
|B_\ell \cdot dA|$, where the integral is over the stellar surface. The equivalent polar 
field strength comes from the axisymmetric configuration for a given spherical harmonic 
degree that yields the same total magnetic flux as that calculated from both the 
axisymmetric and nonaxisymmetric components of the ZDI map. There is a simple analytical 
relation between the equivalent polar field strength and the magnetic flux from ZDI for 
each spherical harmonic degree:
 \begin{equation} 
B_{\rm d} = \frac{1}{2\pi R_\star^2} \Phi_{\rm d},\ 
B_{\rm q} = \frac{3\sqrt{3}}{8\pi R_\star^2} \Phi_{\rm q},\ 
B_{\rm o} = \frac{10}{13\pi R_\star^2} \Phi_{\rm o} \\
 \end{equation}
where $R_\star$ is the stellar radius and $\Phi_{\rm d,q,o}$ is the magnetic flux 
integrated over the surface for the dipole, quadrupole, and octupole components of the 
ZDI map, respectively. The values of $B_{\rm d}, B_{\rm q}, B_{\rm o}$ from our ZDI map 
are listed in Table~\ref{tab1}.

\subsection{X-Ray Data}\label{sec2.2} 

\cite{Poppenhager2009} described previous \xray observations of \Peg from ROSAT, 
XMM-Newton, and Chandra. The ROSAT observation was carried out in late-1992, with a 12~ks 
effective exposure by the PSPC instrument. Sixteen years later in mid-2008, XMM-Newton 
obtained a deep 55~ks pointing on \Peg. Six months after that, Chandra obtained two short 
4.9~ks exposures on the same day with the HRC-I and ACIS-S instruments. 
\citeauthor{Poppenhager2009}\ carried out a detailed analysis of these observations, but 
they adopted a single-temperature plasma model to calculate the crucial energy conversion 
factors that translate count rates into physical fluxes. They obtained a large range of 
\xray luminosities for \Peg, $\log L_X = 26.1$--27.2~erg~s$^{-1}$, with the lower and 
upper limits corresponding to the XMM/MOS1+2 and Chandra/HRC-I measurements, 
respectively.

We carried out an independent assessment of the archival \xray data, using a new modeling 
approach that circumvents the issue of choosing an appropriate single coronal temperature 
\citep{Ayres2022}. We reached similar conclusions to \cite{Poppenhager2009} for the ROSAT 
and Chandra observations. However, as noted by those authors, the deep XMM-Newton 
pointing on \Peg yielded surprisingly minimal detections---probably owing to the use of 
the thick optical blocking filter, which significantly degrades the soft response of the 
camera system. The large energy conversion factors for the XMM pn and MOS modules under 
these circumstances render any derived fluxes problematic. From the ROSAT and Chandra 
data alone, we obtained a range of \xray luminosities $\log L_X = 
26.6$--27.0~erg~s$^{-1}$ in the \mbox{0.1--2.4}~keV band. Considering the minimal 
chromospheric variability of \Peg \citep{Baum2022}, the dispersion in our \xray 
luminosity estimates probably arises from uncertainties in the instrumental calibrations 
at the soft energies that are characteristic of low-activity coronal sources. The 
importance of this calibration issue is evidenced by the fact that the Chandra HRC-I and 
ACIS-S pointings within less than 2\,h of each other yield \xray flux estimates that 
differ by 60\%.

 \begin{deluxetable}{lcc}
  \setlength{\tabcolsep}{14pt}
  \tablecaption{Properties of the Exoplanet Host Star \Peg\label{tab1}}
  \tablehead{\colhead{}            & \colhead{\Peg} & \colhead{Source}}
  \startdata
  $T_{\rm eff}$ (K)                & $5758 \pm 78$             & 1 \\
  $[$M/H$]$ (dex)                  & $+0.18 \pm 0.07$          & 1 \\
  $\log g$ (dex)                   & $4.32 \pm 0.08$           & 1 \\
  $v \sin i$ (km s$^{-1}$)         & $2.0 \pm 0.5$             & 1 \\
  $B-V$ (mag)                      & $0.67$                    & 2 \\
  $\log R'_{\rm HK}$ (dex)         & $-5.068$                  & 2 \\
  $P_{\rm rot}$ (days)             & $21.9 \pm 0.4$            & 3 \\
  $|B_{\rm d}|$ (G)                & $0.770$                   & 4 \\
  $|B_{\rm q}|$ (G)                & $0.441$                   & 4 \\
  $|B_{\rm o}|$ (G)                & $0.652$                   & 4 \\
  $\log L_X$ (erg~s$^{-1}$)        & $26.8 \pm 0.2$            & 5 \\
  Mass-loss rate ($\dot{M}_\odot$) & $0.38 \pm 0.13$           & 5 \\
  $\Delta\nu$ ($\mu$Hz)            & $114.6 \pm 1.2$           & 6 \\
  $\nu_{\rm max}$ ($\mu$Hz)        & $2474 \pm 123$            & 6 \\
  Luminosity ($L_\odot$)           & $1.398 \pm 0.016$         & 6 \\
  Radius ($R_\odot$)               & $1.152 \pm 0.009$         & 6 \\
  Mass ($M_\odot$)                 & $1.09 \pm 0.02$           & 6 \\
  Age (Gyr)                        & $4.8^{+0.7}_{-0.4}$       & 6 \\
  \hline
  Torque ($10^{30}$~erg)           & $0.224^{+0.039}_{-0.075}$ & 7 \\
  \enddata
  \tablerefs{(1)~\cite{Brewer2016}; (2)~\cite{Henry2000}; (3)~\cite{Simpson2010}; 
    (4)~Section~\ref{sec2.1}; (5)~Section~\ref{sec2.2}; (6)~Section~\ref{sec2.3}; 
    (7)~Section~\ref{sec3}}
 \vspace*{-24pt}
 \end{deluxetable}
 \vspace*{-12pt}

Using the empirical relation $\dot{M}\propto F_X^{0.77}$ from \cite{Wood2021}, we can 
estimate the mass-loss rate for \Peg from the range of \xray luminosities determined 
above and the asteroseismic radius in Section~\ref{sec2.3}. The lower bound on the \xray 
luminosity yields $\dot{M} = 0.25\ \dot{M}_\odot$ while the upper bound yields $\dot{M} = 
0.51\ \dot{M}_\odot$. For the estimates of wind braking torque in Section~\ref{sec3}, we 
adopt the average of these two values with an uncertainty that reflects the full range of 
possible \xray luminosities (see Table~\ref{tab1}).

\subsection{Asteroseismology}\label{sec2.3} 

The TESS mission observed \Peg at 20\,s cadence during Sector~56 (2022 September 1--30). 
Following the procedures described in \cite{Metcalfe2023b}, we extracted a custom light 
curve from the TESS target pixel files, substantially improving the signal to noise ratio 
(S/N) compared to the data product from the Science Processing Operations Center. 
Briefly, this procedure starts with a light curve from the pixel with the most signal, 
and includes additional pixels one at a time until the S/N no longer improves. The 
resulting light curve was detrended against centroid pixel coordinates and high-pass 
filtered with a cutoff frequency of 100~$\mu$Hz to remove any residual instrumental 
signatures.

 \begin{figure}[t]
 \centering\includegraphics[width=\columnwidth,trim=10 10 10 10,clip]{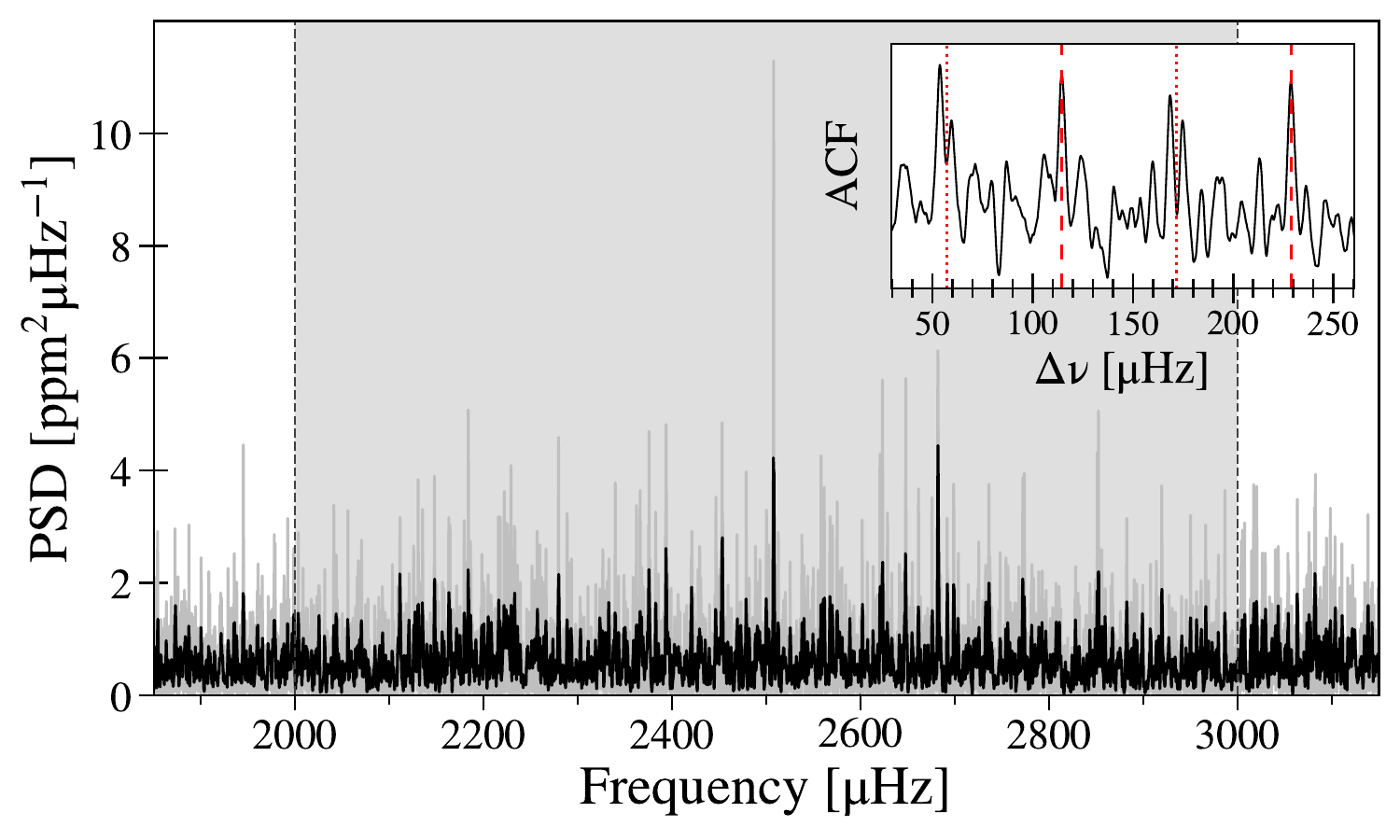}
 \caption{Power spectrum of \Peg centered on the power excess due to solar-like 
oscillations. The shaded region is used to calculate an autocorrelation, shown in the 
inset. Dashed lines in the inset represent expected peaks in the ACF due to the 
characteristic spacings of p-modes.\label{fig3}}
 \end{figure}

Figure~\ref{fig3} shows the power spectrum of \Peg computed from the TESS photometry, 
centered on the power excess near 2500~$\mu$Hz. To confirm that the observed power excess 
is due to solar-like oscillations we used \texttt{pySYD} \citep{Chontos2022, Huber2009}, 
which implements an autocorrelation technique to identify and characterize the global 
oscillation parameters ($\Delta\nu$, $\nu_{\rm max}$). The autocorrelation function (ACF) 
in the inset of Figure~\ref{fig3} is calculated from the shaded region of the power 
spectrum, showing strong peaks at the expected spacings ($\Delta\nu = 114.6 \pm 1.2\ 
\mu$Hz) and confirming an asteroseismic detection with $\nu_{\rm max} = 2474 \pm 123\ 
\mu$Hz.

In addition to the global oscillation parameters, we adopted observational constraints on 
the effective temperature $T_{\rm eff}$ and metallicity [M/H] from \cite{Brewer2016}, as 
well as a bolometric luminosity $L$ derived from the spectral energy distribution 
following the procedures described in \cite{Stassun2017, Stassun2018}. These constraints 
provided the inputs for grid-based modeling with the Yale-Birmingham pipeline 
\citep{Basu2010, Basu2012, Gai2011}, using the same grid of models constructed with YREC 
\citep{Demarque2008} and following the same procedures described in \cite{Metcalfe2021}. 
The resulting determinations of the asteroseismic radius, mass, and age of \Peg are 
listed in Table~\ref{tab1}.

\section{Wind Braking Torque}\label{sec3}

We now have all of the required inputs to estimate the wind braking torque for \Peg, 
following the prescription of 
\cite{FinleyMatt2018}\footnote{\url{https://github.com/travismetcalfe/FinleyMatt2018}}. 
Bringing together the equivalent polar field strengths from our ZDI map in 
Section~\ref{sec2.1}, the mass-loss rate from the empirical relation of \cite{Wood2021} 
in Section~\ref{sec2.2}, the chromospheric rotation period from \cite{Simpson2010}, and 
the asteroseismic mass and radius from Section~\ref{sec2.3}, we calculate a wind braking 
torque of $0.224^{+0.039}_{-0.075} \times 10^{30}$~erg. The uncertainty includes a 
contribution from the inclination, evaluated at the extremes of the 68\% credible 
interval between $42^\circ$ and $78^\circ$. For each inclination, we inverted a new ZDI 
map from the observations, calculated the equivalent polar field strengths for each 
spherical harmonic degree, and updated our estimate of the wind braking torque. With the 
other parameters fixed, the resulting torque at both extremes of the inclination was 
slightly lower than for our fiducial model (0.196 and 0.214$\times 10^{30}$~erg at 
$42^\circ$ and $78^\circ$, respectively), bolstering our conclusions.

\begin{figure}[t]
 \centering\includegraphics[width=\columnwidth]{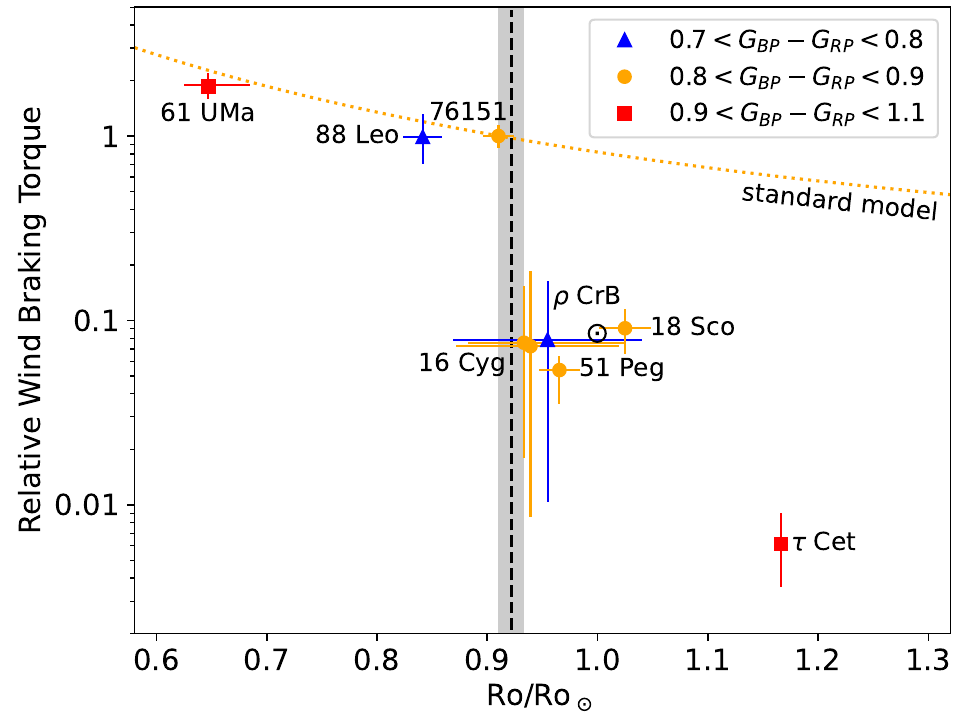}
 \caption{Estimated wind braking torque relative to \HD as a function of Rossby number 
normalized to the solar value. Points are grouped by Gaia color, corresponding to solar 
analogs (yellow circles) and hotter (blue triangles) or cooler (red squares) stars. The 
gray shaded area represents our empirical constraint on the critical Rossby number for 
the onset of WMB ($\mathrm{Ro_{crit}}/\mathrm{Ro_\odot}=0.92\pm0.01$).\label{fig4}}
 \end{figure}

In Figure~\ref{fig4}, we compare \Peg with similarly estimated wind braking torques for 
two slightly hotter stars \citep{Metcalfe2021}, four solar analogs \citep{Metcalfe2022}, 
and two cooler G-type stars \citep{Metcalfe2023a}. Rossby numbers were calculated from 
the Gaia $G_{BP}-G_{RP}$ color using the asteroseismic calibration of \cite{Corsaro2021}, 
normalized to the solar value on this scale (Ro$_\odot=0.496$). The wind braking torque 
has been normalized to the value for \HD ($4.17 \times 10^{30}$~erg) to facilitate a 
comparison with theoretical models. Horizontal error bars come from the uncertainty in 
the rotation period, while vertical error bars reflect the range of possible torques when 
all input quantities are shifted by $\pm1\sigma$. The gray shaded area represents our 
empirical constraint on the critical Rossby number for the onset of WMB 
($\mathrm{Ro_{crit}} / \mathrm{Ro_\odot}=0.92 \pm 0.01$), and the dotted yellow line 
shows the evolution of the torque for \HD ($M=1.05\ M_\odot$) from a standard spin-down 
model \citep{vanSaders2013}. The mass-dependence of stellar spin-down can shift this 
standard model up or down by a factor of two for the mass range shown in 
Figure~\ref{fig4}.

The wind braking torque for \Peg clearly places it in the WMB regime, with the ZDI map 
providing a much stronger constraint than $\rho$\,CrB or the components of \Cyg, which 
rely on upper limits from statistical non-detections of the large-scale magnetic field. 
We compare the fiducial models of \Peg and \HD \citep[see][]{Metcalfe2022} to evaluate 
the relative importance of various contributions to the total decrease in torque. The 
wind braking torque decreases by nearly a factor of 20 ($-$95\%) between the ages of 
these two stars (2.6--4.8 Gyr), dominated by changes in the mass-loss rate ($-$81\%) and 
magnetic field strength and morphology ($-$78\%) with smaller contributions from the 
differences in rotation period ($-$6\%) and stellar mass ($-$0.9\%). These decreases are 
substantially offset by evolutionary changes in the stellar radius ($+$75\%). The overall 
decrease in the wind braking torque is larger for \Peg (by 4.8~Gyr) than from a similar 
comparison of \Sco (at 3.7~Gyr) and \HD in \cite{Metcalfe2022}, but it reinforces the 
relatively equal importance of changes in the mass-loss rate and magnetic field strength 
and morphology ($-$69\% and $-$65\%, respectively for \Sco) near the onset of WMB.

A comparison of the fiducial models for \Peg and \Sco suggests that the wind braking 
torque continues to decrease ($-$41\%) at later evolutionary phases (3.7--4.8~Gyr). As 
with the components of \Cyg \citep[see][]{Metcalfe2022}, the subsequent decrease in wind 
braking torque becomes dominated by the evolution of magnetic field strength and 
morphology ($-$40\%) with smaller contributions from differences in the mass-loss rate 
($-$36\%) and stellar mass ($-$2\%), offset by evolutionary changes in stellar radius 
(+51\%) and the difference in rotation period (+4\%). Although changes in the magnetic 
field strength and morphology continue to favor a lower wind braking torque ($-$33\%) 
between the ages of \Peg and \CygA (4.8--7.0~Gyr), this decrease is overwhelmed by the 
evolution of mass-loss rate (+61\%) and stellar radius (+21\%), with small contributions 
from differences in rotation period (+7\%) and stellar mass (+0.4\%).

Standard spin-down models fail to predict the substantial changes in wind braking torque 
that are suggested by the observations. While these models generally reproduce the 
evolution prior to the onset of WMB (dotted line in Figure~\ref{fig4}), they predict a 
decrease of only $-$12\% in the wind braking torque between \HD and \Peg, rather than the 
$-$95\% estimated above. This failure can be traced to underestimated changes in both the 
mass-loss rate (or $L_X$) and the magnetic field strength, as well as neglected changes 
in magnetic morphology. Standard models scale the mass-loss rate as $\dot{M} \sim L / 
\mathrm{Ro}^{2}$ and the magnetic field strength as $B \sim P_{\rm phot}^{1/2} / 
\mathrm{Ro}$, where $P_{\rm phot}$ is the photospheric pressure \citep{vanSaders2013}. 
According to these models, the mass-loss rate is predicted to decrease by $-9\%$ and the 
magnetic field strength by $-$6\% between \HD and \Peg, while the observations suggest a 
decrease of $-$95\% in the mass-loss rate and $-$57\% in the magnetic field strength 
(from the difference in the activity proxy $\log R'_{\rm HK}$). The average strength of 
the large-scale field from spectropolarimetry decreases by $-$78\% \citep{See2019}, 
suggesting that changes in magnetic morphology are also important.

\section{Discussion}\label{sec4}

Using new observational constraints from LBT spectropolarimetry and TESS 
asteroseismology, we have demonstrated that the wind braking torque of the exoplanet host 
star \Peg places it firmly in the WMB regime. This provides a natural explanation for the 
disagreement between the observed rotation period (21.9~days) and that predicted from its 
mean activity level \citep[29~days;][]{Wright2004}. At the onset of WMB, rotation and 
activity decouple \citep{Metcalfe2016, Metcalfe2019a} as the magnetic fields that had 
previously facilitated the efficient loss of angular momentum to stellar winds grow 
weaker and cascade to smaller spatial scales. Beyond this transition, the rotation period 
only changes with the stellar moment of inertia \citep{vanSaders2016} while the activity 
level continues to decline with age \citep{LorenzoOliveira2018, Huber2022} as mechanical 
energy from convection becomes the dominant source of chromospheric heating 
\citep{BohmVitense2007}. Our asteroseismic age for \Peg is older than expected from its 
rotation period \citep[$2.9\pm0.5$~Gyr;][]{Barnes2007}, but younger than expected from 
its activity level \citep[7~Gyr;][]{Donahue1998}, so other factors such as non-solar 
metallicity may also contribute to the erroneous prediction of rotation period from 
activity level \citep[see][]{SaarTesta2012}. It is now clear that WMB begins before stars 
reach Ro$_\odot$, and our empirical constraint on the value of 
$\mathrm{Ro_{crit}}/\mathrm{Ro_\odot}$ is consistent with that derived from the analysis 
of a larger sample of stars with asteroseismic rotation periods and ages 
\citep{Saunders2023}.

Our revised stellar properties and estimated wind braking torque for \Peg have minimal 
consequences for the characterization of the exoplanet, but interesting implications for 
the current space weather environment of the system. The most recent orbital solution for 
\Pegb \citep{Rosenthal2021} adopted a stellar mass that was only 2\% lower than the 
asteroseismic mass determined in Section~\ref{sec2.3}. The resulting update to the 
planetary mass would be well within the quoted uncertainties. Considering our 
characterization of the large-scale field and mass-loss rate of \Peg, direct magnetic 
interactions between the star and planet are unlikely. The Alfv\'en radius of our 
fiducial wind model for \Peg is $R_A = 4.7\ R_\star$, while the semi-major axis of \Pegb 
is much larger at $a = 9.8\ R_\star$. Even prior to the onset of WMB, assuming the 
magnetic field properties and mass-loss rate of \HD \citep{Metcalfe2022}, the Alfv\'en 
radius of \Peg would have been $R_A = 5.5\ R_\star$, still well inside the planetary 
orbit. Nevertheless, standard spin-down models predict that without WMB, \Peg would have 
had both a higher mass-loss rate and a stronger magnetic field with more large-scale open 
field where energetic eruptions could escape \citep{Garraffo2015}, creating a harsher 
space weather environment than actually exists. Consequently, older stars beyond the 
onset of WMB may provide a more stable environment for the development of technological 
civilizations.

\vspace*{12pt}
T.S.M.\ is supported by NASA grant 80NSSC22K0475 and NSF grant AST-2205919.
D.B.\ gratefully acknowledges support from NASA (NNX16AB76G, 80NSSC22K0622) and the Whitaker Endowed Fund at Florida Gulf Coast University.
O.K.\ acknowledges support by the Swedish Research Council (grant agreement no.\ 2019-03548), the Swedish National Space Agency, and the Royal Swedish Academy of Sciences.
S.B.\ is supported by NSF grant AST-2205026.
V.S.\ acknowledges support from the European Space Agency (ESA) as an ESA Research Fellow.
A.G.S.\ acknowledges support from the National Science Foundation Graduate Research Fellowship Program under Grant No.~1842402.
The LBT is an international collaboration among institutions in the United States, Italy and Germany. 
This paper includes data collected with the TESS mission, obtained from the Mikulski Archive for Space Telescopes at the Space Telescope Science Institute (STScI). The specific observations analyzed can be accessed via \dataset[doi:10.17909/emgj-gq45]{https://doi.org/10.17909/emgj-gq45}. 
Funding for the TESS mission is provided by the NASA Explorer Program. STScI is operated by the Association of Universities for Research in Astronomy, Inc., under NASA contract NAS 5-26555.
This research was supported in part by the Nonprofit Adopt a Star program (\href{https://adoptastar.org}{\mbox{adoptastar.org}}) administered by White Dwarf Research Corporation.


\end{document}